\documentclass[prb,twocolumn,groupedaddress,showpacs,aps,floatfix,fleqn]{revtex4}

\usepackage{amsmath}
\usepackage{amstext}
\usepackage{epsfig}
\usepackage{graphicx}
\usepackage{dcolumn}
\usepackage{rotating}
\usepackage{draftcopy}
\usepackage{flafter}
\usepackage[below]{placeins}
\usepackage{ccaption}
\usepackage{soul}
\usepackage{wrapfig}
\usepackage{sidecap}
\usepackage{float}
\usepackage{subfig}
\usepackage{booktabs}
\usepackage{tabularx}

\graphicspath{{Figs/}}

\begin{document}
\bibliographystyle{apsrev}
\title{Optical Dielectric Functions of III-V Semiconductors in Wurtzite Phase}

\author{Amrit De}
\author{Craig E. Pryor}

\affiliation{Department of Physics and Astronomy and Optical Science and Technology Center, \\ University of Iowa, Iowa City, Iowa 52242}

\date{\today}

\begin{abstract}
Optical properties of semiconductors can exhibit strong polarization dependence due to crystalline anisotropy. A number of recent experiments have shown that the photoluminescence intensity in free standing nanowires is polarization dependent.
One contribution to this effect is the anisotropy of the dielectric function due to the fact that most nanowires crystalize in the wurtzite form.
While little is known experimentally about the band structures wurtzite phase III-V semiconductors, we have previously predicted the bulk band structure of nine III-V semiconductors in wurtzite phase.
Here, we predict the frequency dependent dielectric functions for nine non-Nitride wurtzite phase III-V semiconductors (AlP, AlAs, AlSb, GaP, GaAs, GaSb, InP, InAs and InSb). Their complex dielectric functions are calculated in the dipole approximation by evaluating the momentum matrix elements on a dense grid of special k-points using empirical pseudopotential wave functions. Corrections to the momentum matrix elements accounting for the missing core states are made using a scaling factor which is determined by using the optical sum rules on the calculated dielectric functions for the zincblende polytypes.
The dielectric function is calculated for polarizations perpendicular and parallel to the $c$-axis of the crystal.
\end{abstract}

\pacs{
71.15.Dx, % Computational methodology (Brillouin zone sampling, iterative diagonalization, pseudopotential construction)
71.20.-b, %Electron density of states and band structure of crystalline solids
71.00.00 % Electronic structure of bulk materials
}
\maketitle

%------------------- INTRODUCTION ---------------------------
\section{Introduction}
Frequency-dependent reflectivity and absorption spectra have been a fundamental tool for studying the band structures of semiconductors. In semiconductors with wurtzite (WZ) structure, the optical properties are very strongly polarization dependent \cite{Cardona1965,Reynolds1965,Xu1993,Ninomiya1995,Kawashima1997,Alemu1998} due to the crystalline anisotropy. The optical selection rules for WZ \cite{Birman1959b,Streitwolf1969} dictate which polarization dependent interband transitions are allowed. This helps in further resolving the band structure for WZ and is used to determine crystal-field splitting and spin-orbit splitting energies.

A number recent experiments have measured the optical polarization dependent photoluminescence(PL) intensity in free standing nanowires \cite{Wang2001,Mattila2006,Mattila2007,Mishra2007,Lan2008,Kobayashi2008,Novikov2010}. These results reveal a very strong polarization dependence in the PL intensity depending on whether the incident light is polarized parallel and perpendicular to the growth axis of the nanowire. This polarization anisotropy has been attributed to effects such as heavy and light hole mixing\cite{Sercel:1990p646}, large dielectric contrast between the nanowires and its surroundings\cite{Wang2001} and intrinsic band structure properties of the nanowire\cite{Persson2004}.

However, it is well known that the nanowires tend to crystalize in WZ phase rather than zincblende(ZB) (which is the more stable phase for non-Nitride bulk III-Vs). The polarization dependent optical anisotropy of the nanowires will have a contribution from the WZ crystal structure.
%However very little is known about the WZ phase of III-V semiconductors as they do not naturally occur.
% this paper predicts the optical properties, so it *is* possible
% Hence it is not possible predict their optical properties or interpret reflectivity measurements.
% replace the above line with this:
However, without bulk samples it is not possible to determine the optical properties with traditional reflectivity measurements.

In Ref. \onlinecite{De2010prb} we predicted the bulk band structure of nine non-Nitride III-V semiconductors in WZ phase using empirical pseudopotentials, including SO interactions. These calculations were based on transferable model pseudopotentials assuming ideal relations between the ZB and WZ phases and their lattice constants. The spherically symmetric ionic model potentials for the ZB phase were first obtained by fitting the calculated transition energies to experimental transition energies at high symmetry points. The WZ phase band structures were then obtained by transferring the model ionic pseudopotentials to the WZ pseudopotential Hamiltonian. This is justified because in both ZB and WZ, all of the nearest neighbors and nine out of the twelve second nearest neighbors are at identical crystallographic locations \cite{Birman.1959}, while all the second nearest neighbors are equidistant. This method has proven to be quite successful in the past, in obtaining band structures of polytypes \cite{Joannopoulos1973,Pugh1999,Pennington2001,Fritsch2006,Cohen.book}. Our agreement with experiment was excellent for the known band gaps of InP, InAs and GaAs.

In this article, we predict the frequency dependent linear dielectric function for the WZ phase of nine III-V semiconductors, based on our previous empirical pseudopotential band structure calculations. The dielectric functions are calculated in the linear optical response regime within the electric dipole approximation. The required momentum matrix elements are obtained by using the calculated wave functions from our empirical pseudopotentials\cite{De2010prb}. Momentum matrix elements must be corrected because the pseudo-wave functions do not include the core states. Such corrections have been made using nonlocal correction terms \cite{Kageshima1997,Adolph2001,Pickard2001,Monachesi2001}. Since the core states of the two polytypes should be very nearly the same, such core corrections should be the same for ZB and WZ forms. Therefore, we use the optical sum rules to obtain a set of normalization constants which yield $\epsilon_0\equiv\epsilon(\omega = 0)$ for the ZB phases that fit experimental values. These normalization constants are then transferred to their respective WZ polytypes to obtain their dielectric functions.

Our paper is organized as follows. In Sec. \ref{sec:optical}, a brief theoretical background of the optical properties is presented along with our method for the calculations. The calculated dielectric functions and reflectivity spectra are presented in Sec. \ref{sec:results} along with our tabulated results for $\epsilon_0$. This is followed by a brief discussion and summary.

%=============================================================================================================================
\section{Optical Properties\label{sec:optical}}
%========================================================================================================

Consider a semi-infinite crystal having symmetry equivalent to or higher than that of the orthorhombic crystal system. If we choose our coordinate such that the $z$-axis is the surface normal and the $x$-$z$ plane is the plane of incidence, then the reflectivity for light polarized perpendicular to the plane of incidence is given by \cite{Mosteller1968,Brehat1991}

\begin{equation}
R_s=\left|\frac{\cos\theta-(n^2_y-\sin^2{\theta})^{\frac{1}{2}}}{\cos\theta+(n^2_y-\sin^2{\theta})^{\frac{1}{2}}}\right|^2
\label{Rs}
\end{equation}
where $\theta$ at an angle of incidence. Similarly the reflectivity for light polarized parallel to the plane of incidence is
\begin{equation}
R_p=\left|\frac{n_xn_z\cos\theta-(n^2_z-\sin^2{\theta})^{\frac{1}{2}}}{n_xn_z\cos\theta+(n^2_z-\sin^2{\theta})^{\frac{1}{2}}}\right|^2
\label{Rp}
\end{equation}

\noindent where $n_i$ ($i=x,y$ or $z$) is the complex index of refraction. In case of an optically uniaxial crystal such as WZ, if the surface normal is parallel to the $c$-axis then $n_z=n_{||}$ and $n_x=n_y=n_{\perp}$. In the linear response regime, $n(\omega)=\sqrt{\epsilon(\omega)}$, which can be written as
\begin{eqnarray}
\epsilon(\omega)=\epsilon^{\prime}(\omega)+i\epsilon^{\prime\prime}(\omega).
\label{eps}
\end{eqnarray}

\noindent In the linear response regime, the real and imaginary parts of the dielectric function are not independent of each other, but obey the Kramer-Kronig (KK) relations

\begin{eqnarray}
\epsilon^{\prime}(\omega)=1+\frac{2}{\pi} {\mathcal P}\displaystyle\int_0^\infty\frac{\omega'\epsilon^{\prime\prime}(\omega)}{\omega'^2-\omega^2}d\omega'
\label{eps1_KK}
\end{eqnarray}

\begin{eqnarray}
\epsilon^{\prime\prime}(\omega)=-\frac{2}{\pi} {\mathcal P}\displaystyle\int_0^\infty\frac{\omega\epsilon^{\prime}(\omega)}{\omega'^2-\omega^2}d\omega'
\label{eps2_KK}
\end{eqnarray}\vspace{1 mm}
where $ {\mathcal P}$ is the Cauchy principle value.

To calculate the optical properties of the WZ semiconductors, we first evaluate $\epsilon^{\prime\prime}(\omega)$ based on our earlier empirical pseudopotential band structure calculations\cite{De2010prb} and then obtain $\epsilon^{\prime}(\omega)$ using the KK relations.
In the electric dipole approximation, assuming only direct band to band transitions are allowed between an initial state $i$ and a final state, $j$, $\epsilon^{\prime\prime}(\omega)$ is given by
\begin{eqnarray}
\epsilon^{\prime\prime}(\omega)&=& \left(\frac{\hbar\pi^2e^2}{m^2\omega^2}\right)\times\\\nonumber
 &~&\displaystyle\sum_{ij}\displaystyle{\int_{BZ}}|M_{ij}|^2\delta(E_{c,j}({\bf k})-E_{v,i}({\bf k})-\hbar\omega)d^3k
\label{eps2}
\end{eqnarray}\vspace*{1 mm}
where $\int_{BZ}$ is an integration over the entire Brillouin zone (BZ), $\sum_{ij}$ is a sum over all initial valance band and final conduction band states, and $E_v(\bf k)$ and $E_c(\bf k)$ are the valance and conduction band energies at their respective $\bf k$s.
%For intrinsic semiconductors, the Fermi level is lies in between the valance and conduction band states.
% The indices $i$ and $j$ are therefore be assigned accordingly.
For the delta function, we use
\begin{eqnarray}
\delta(\Delta E -\hbar\omega)\approx\frac{2}{1+ \cosh[\gamma(\Delta E -\hbar\omega)]}.
\label{delta}
\end{eqnarray}
where $\gamma$ is an adjustable damping parameter that can be used to incorporate lifetime broadening effects. We used $\gamma=100~\rm eV^{-1}$ which gives a transition linewidth of about $35~\rm meV$ \cite{Wang1981}.

The momentum matrix elements $M_{ij}$ for interband transitions are obtained from the pseudo-wave functions from our empirical pseudopotentials for WZ\cite{De2010prb}, given by
\begin{eqnarray}
\phi_{\bf k}({\bf r})=\displaystyle\sum_{\bf G}c({\bf k,G})\exp[i{\bf (k+G)}\cdot{\bf r}]
\label{P_psi}
\end{eqnarray}\vspace{1mm}
where $c({\bf k,G})$ are the eigenvector coefficients at a given $\bf k$. The momentum matrix element between the states $i$ and $j$ is
\begin{eqnarray}
M_{ij}({\bf k})=\langle\phi_{\bf k}^i|\hat{p}|\phi_{\bf k}^j\rangle
\label{Mij_raw}
\end{eqnarray}

where $\hat p$ is the momentum operator. Using Eq. (\ref{P_psi}), $M_{ij}$ can be rewritten in terms of the expansion coefficients as,
\begin{eqnarray}
M_{ij}({\bf k})=i\displaystyle{\sum_{\bf G}}c_i^*({\bf k,G})c_j({\bf k,G})[\bf{(k+G)}\cdot{\bf\hat e}]
\label{Mij}
\end{eqnarray}
 where $\bf\hat e$ is the polarization vector. The expansion coefficients, $c_i$ and $c_j$, are the eigenvectors of $i^{th}$ and $j^{th}$ state in the pseudopotential Hamiltonian. $\epsilon^{\prime\prime}(\omega)$ is then calculated using Eqs. \ref{eps2_KK}, \ref{eps2}, \ref{Mij} for light polarized parallel and perpendicular to the $c$-axis. In practice, it is difficult to explicitly evaluate the Brillouin zone integral in Eq.\ref{eps2} because of the prohibitively large number of $k$ values that would be required.
  However, integration schemes that allow the BZ integral to be replaced by a sum over a set of special $k$ points can be used.
We have used a set of $4.5\times 10^4$  special $k$ points based on the scheme of Monkhorst and Pack \cite{Monkhorst1976}.

\begin{figure}
\centering
\includegraphics[width=1.0\columnwidth]{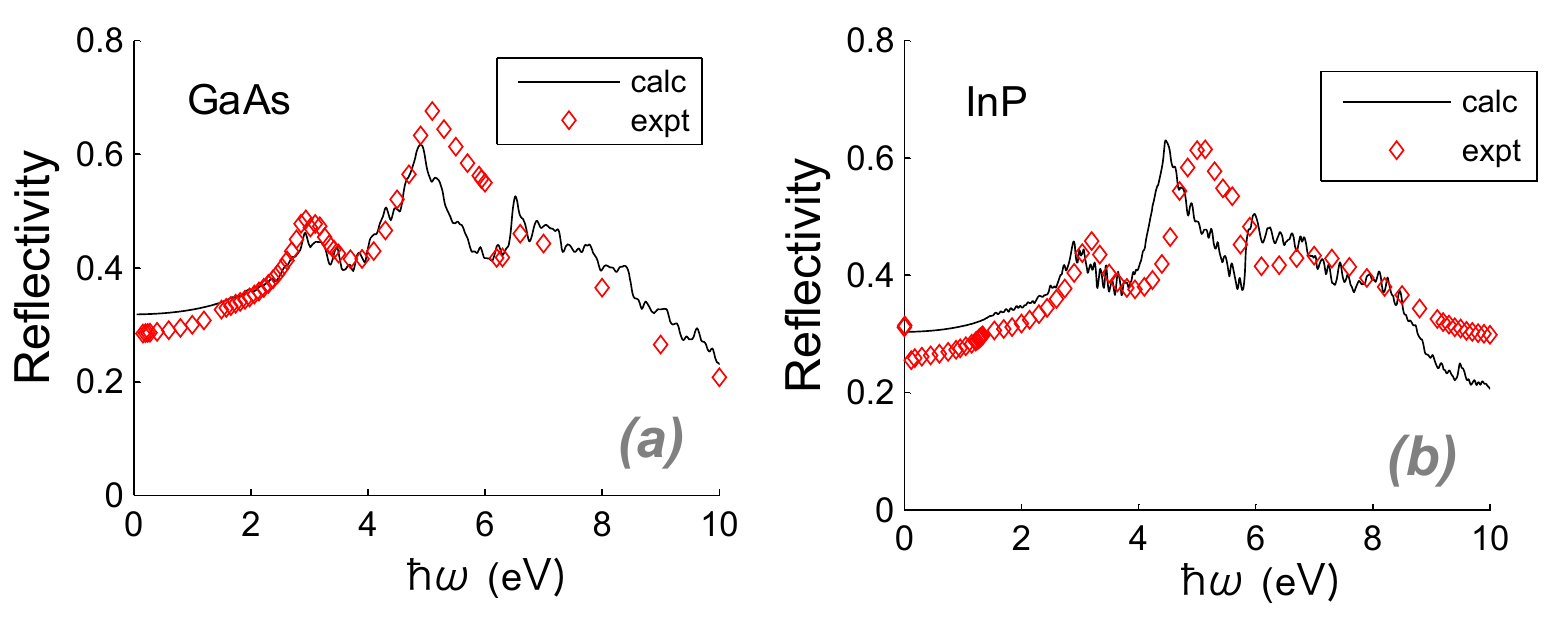}
\caption{ Comparison between calculated and measured reflectivity spectra at normal incidence for zincblende phase {\bf(a)} GaAs and {\bf(b)} InP }
\label{fig:R_ZB}
\end{figure}

%%%%%%%%%%% re-visit
Momentum matrix elements calculated using the pseudo-wave functions must be corrected since the pseudo-wave functions do not include the core states.
One method is to include the commutator of the nonlocal pseudopotential and the position operator \cite{Read1991,Adolph1996}, while Kageshima and Shiraishi have proposed correcting the momentum matrix elements by including a core repair term \cite{Kageshima1997}.
Both techniques cause small changes the dielectric function.
%however the correction typically changes the resulting dielectric function by less than $5\%$.
Monachesi {\it et al.} have compared the dielectric functions of GaAs calculated with {\it ab initio} pseudopotential wave functions against calculations with true electron wave functions and find virtually no differences due to the missing core states\cite{Monachesi2001}.
In our calculations we take advantage of the fact that our pseudopotentials are being transferred between polytypes, and hence any core corrections should be nearly identical.
We normalize the calculated $\epsilon(\omega=0)$ to the experimentally known static dielectric constant by making use of the optical sum rule
\begin{equation}
  \epsilon_o=1+C\frac{2}{\pi}\displaystyle\int_0^\infty\frac{\epsilon^{\prime\prime}(\omega)d\omega}{\omega}
\label{f-sum}
\end{equation}
\noindent where $C$ is a scaling constant which is determined as follows. First the dielectric functions for all the ZB phase semiconductors are evaluated using empirical pseudopotential wave functions. The constant $C$ is then adjusted so that the calculated $\epsilon_0$ for ZB matches experimental values obtained from Ref. \onlinecite{Madelung}. By empirically fitting to $\epsilon _0$ this method results in good agreement between theory and experiment for frequencies $\omega > 0$. As an example, we show a comparison between the calculated and measured reflectivity spectra at normal incidence for cubic GaAs and InP in Fig. \ref{fig:R_ZB}. The measured dielectric functions for GaAs and InP were obtained from Ref. \onlinecite{Palik.Handbook}.

For the WZ calculations, $\epsilon^{\prime\prime}_{\perp}$ and $\epsilon^{\prime\prime}_{||}$ obtained using Eq. \ref{eps2} are multiplied by the same scaling constant, $C$. We expect this method to yield good results for the WZ semiconductors since both polytypes consist of the same atomic species, so the corrections to account for missing core states will be the same.

%=============================================================================================================================
\section{Results and Discussion \label{sec:results}}
%================================================================================================================
The calculated $\epsilon_{0}$ for light polarized parallel and perpendicular to the $c$-axis are listed in table.\ref{tab:eps}. The values listed in the table show that generally, materials with heavier elements have larger static dielectric constants.
We also see that $\epsilon_{0}^{\perp} < \epsilon_{0}^{||}$ in the case of GaAs, GaSb and InSb, whereas  $\epsilon_{0}^{\perp} > \epsilon_{0}^{||}$ for AlP, AlAs, GaP InP, and InAs.

\begin{figure}
\centering
\includegraphics[width=1.0\columnwidth]{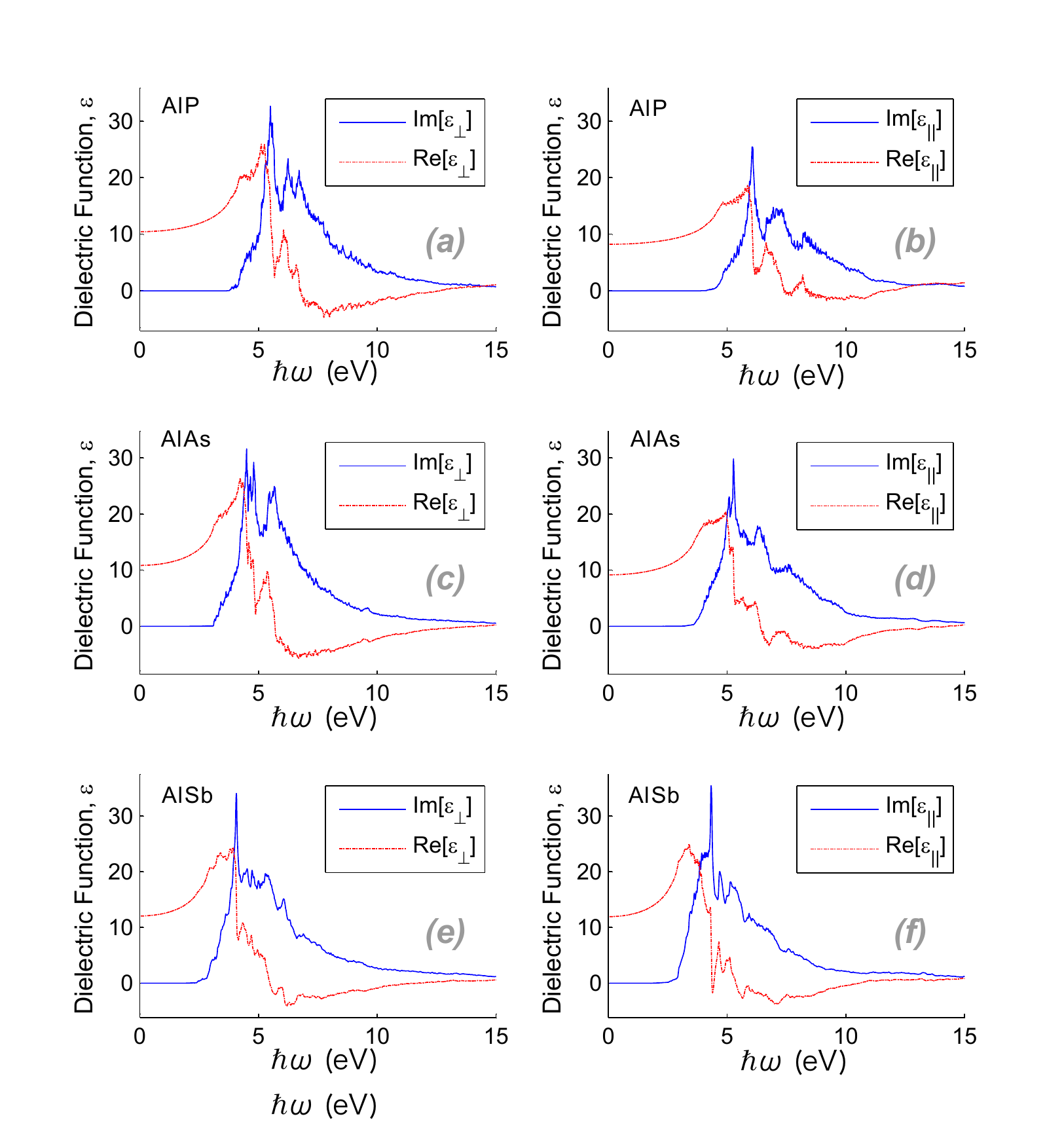}
\caption{Real and imaginary parts of the complex dielectric function as a function of incident photon energies and polarizations for {\bf(a)}$E_{\perp}$ in AlP, {\bf(b)}$E_{||}$ in AlP, {\bf(c)}$E_{\perp}$ in AlAs, {\bf(d)} $E_{||}$ in AlAs, {\bf(e)}$E_{\perp}$ in AlSb and {\bf(f)}$E_{||}$ in AlSb }
\label{fig:eps:AlPAsSb}
\end{figure}

\begin{figure}
\centering
\includegraphics[width=1.0\columnwidth]{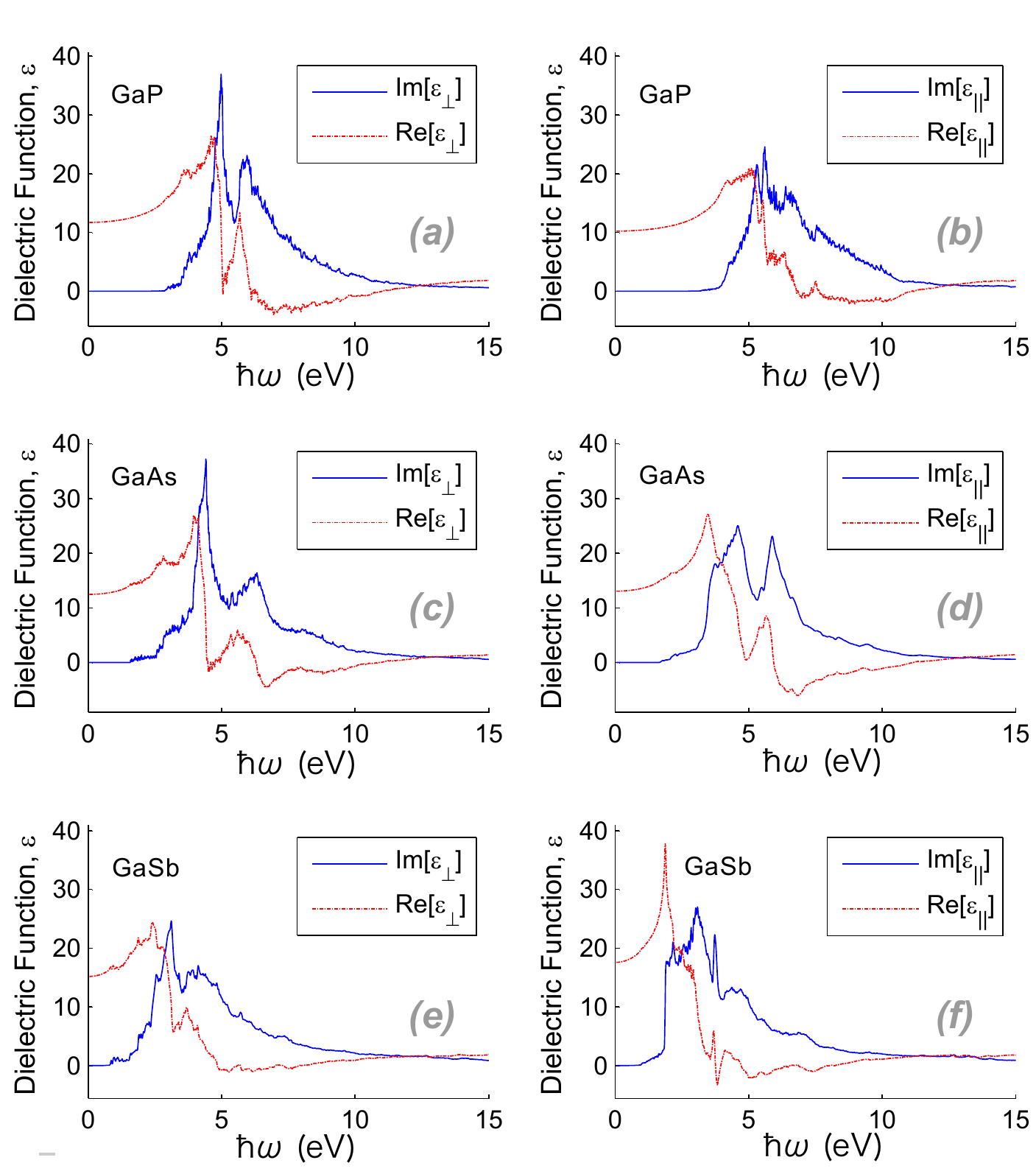}
\caption{ Real and imaginary parts of the complex dielectric function as a function of incident photon energies and polarizations for {\bf(a)}$E_{\perp}$ in GaP, {\bf(b)}$E_{||}$ in GaP, {\bf(c)}$E_{\perp}$ in GaAs, {\bf(d)} $E_{||}$ in GaAs, {\bf(e)}$E_{\perp}$ in GaSb and {\bf(f)}$E_{||}$ in GaSb }
\label{fig:eps:GaPAsSb}
\end{figure}

\begin{figure}
\centering
\includegraphics[width=1.0\columnwidth]{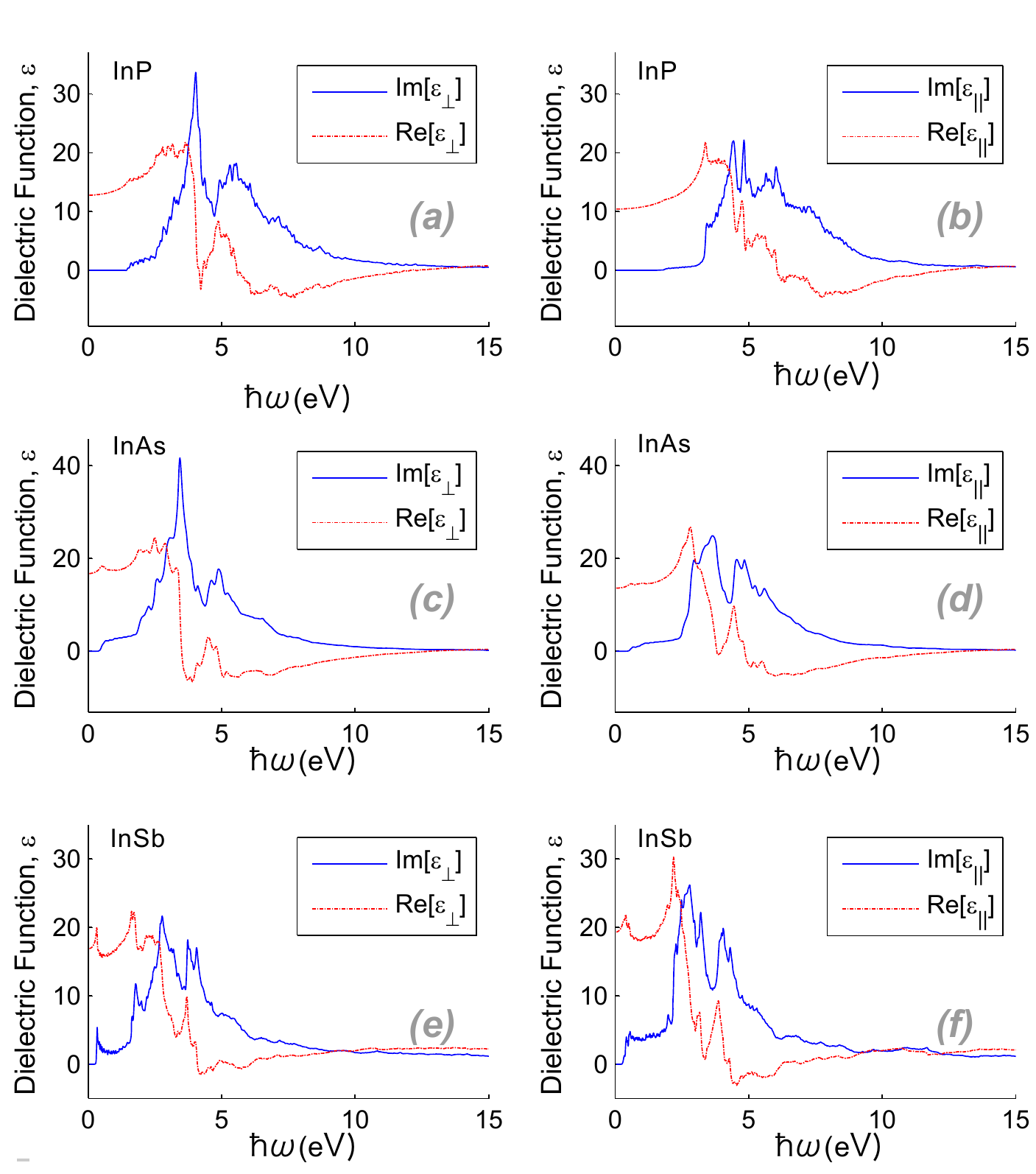}
\caption{ Real and imaginary parts of the complex dielectric function as a function of incident photon energies and polarizations for {\bf(a)}$E_{\perp}$ in InP, {\bf(b)}$E_{||}$ in InP, {\bf(c)}$E_{\perp}$ in InAs, {\bf(d)} $E_{||}$ in InAs, {\bf(e)}$E_{\perp}$ in InSb and {\bf(f)}$E_{||}$ in InSb }
\label{fig:eps:InPAsSb}
\end{figure}

\begin{table}
\begin{center}
\begin{tabular}{c|c c}
%\toprule
 \hline
 \hline
{Material}  & $\epsilon_{0}^{\perp}$ & $\epsilon_{0}^{||}$ \\
\hline
AlP    &  10.464  &  8.232   \\
AlAs   &  10.853  &  9.165   \\
AlSb   &  12.056  & 11.933   \\
GaP    &  11.708  & 10.223   \\
GaAs   &  12.481  & 13.066   \\
GaSb   &  15.215  & 17.621   \\
InP    &  12.812  & 10.435   \\
InAs   &  16.782  & 13.610   \\
InSb   &  16.952  & 19.379   \\
\hline
%\bottomrule
\end{tabular}
\caption {Calculated static dielectric constant for light polarized parallel and perpendicular to the $c$-axis, for nine WZ phase III-V semiconductors. \label{tab:eps} }
\end{center}
\end{table}

\begin{figure}
\centering
\includegraphics[width=0.65\columnwidth]{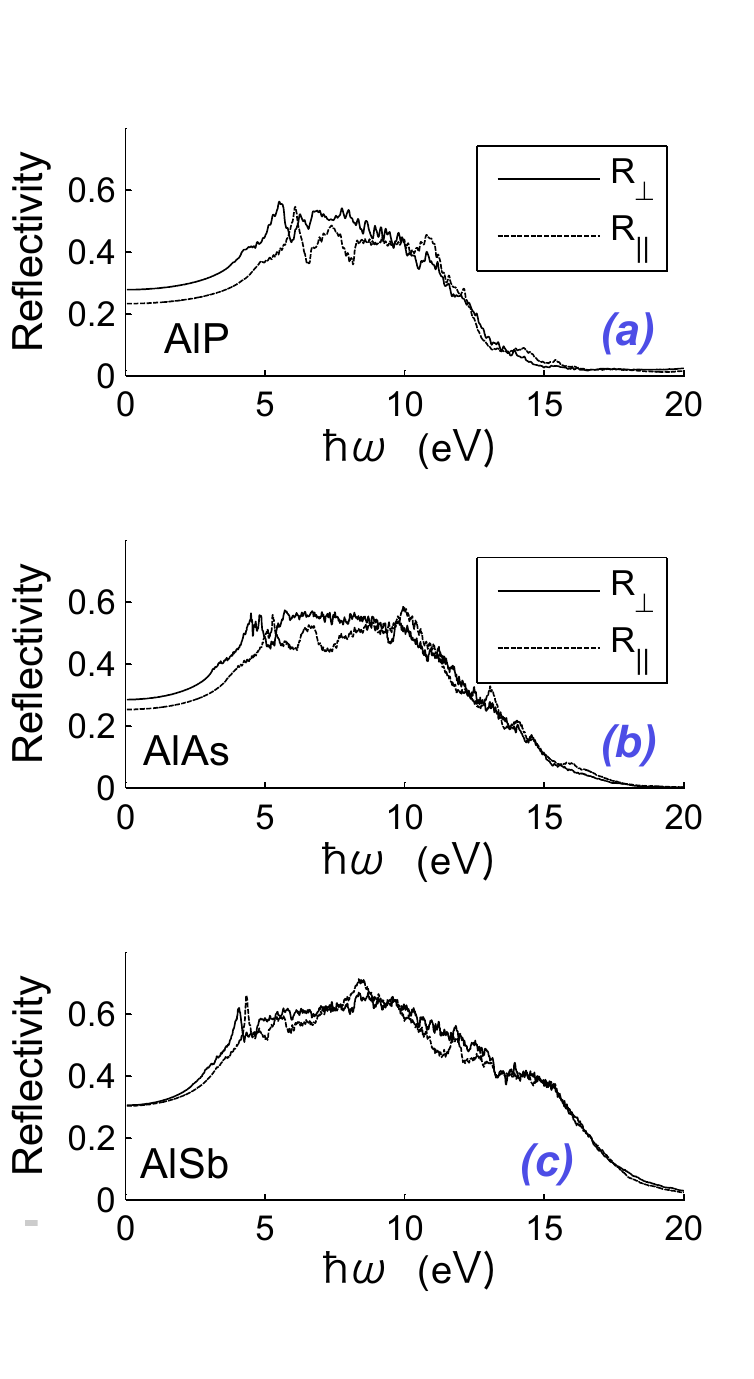}
\caption{ Calculated reflectivity spectra at normal incidence for {\bf(a)} AlP, {\bf(b)} AlAs and {\bf(c)} AlSb }
\label{fig:R:AlPAsSb}
\end{figure}

\begin{figure}
\centering
\includegraphics[width=0.65\columnwidth]{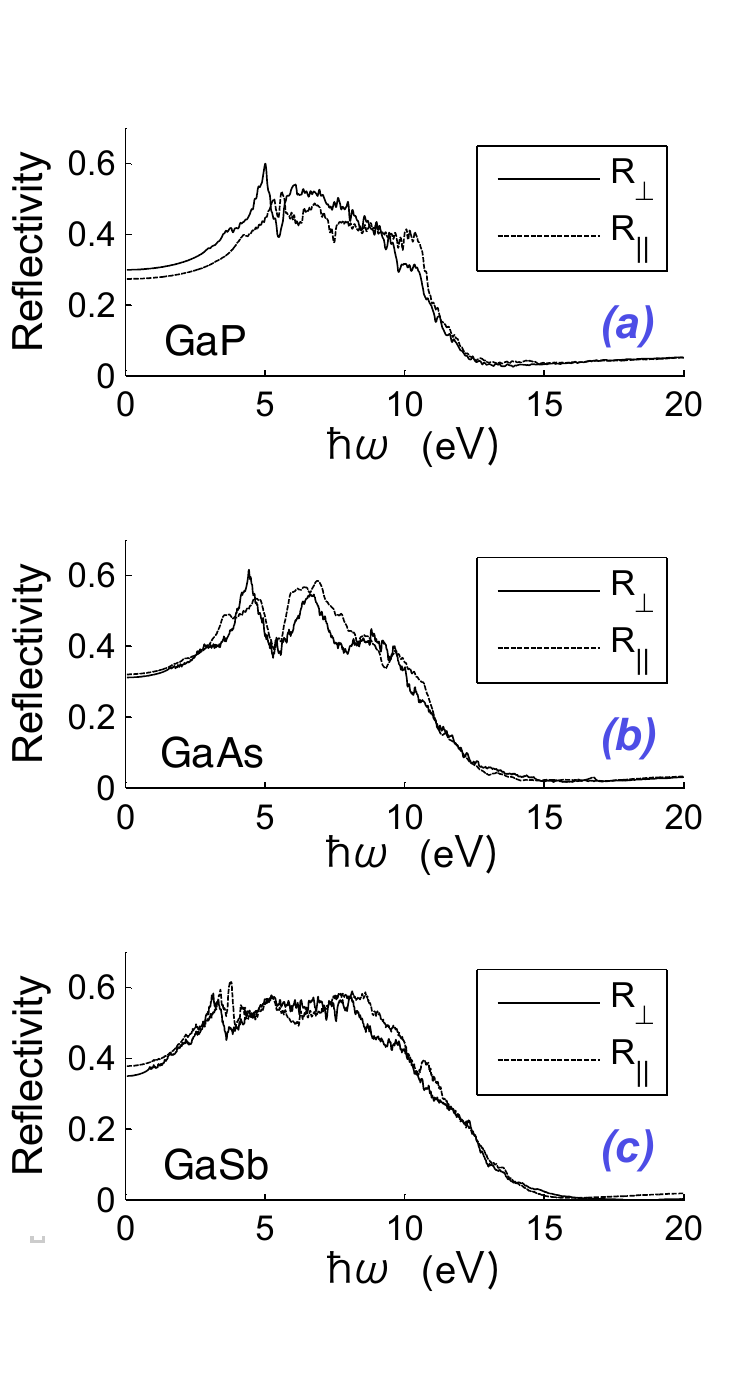}
\caption{ Calculated reflectivity spectra at normal incidence for {\bf(a)} GaP, {\bf(b)} GaAs and {\bf(c)} GaSb }
\label{fig:R:GaPAsSb}
\end{figure}

\begin{figure}
\centering
\includegraphics[width=0.65\columnwidth]{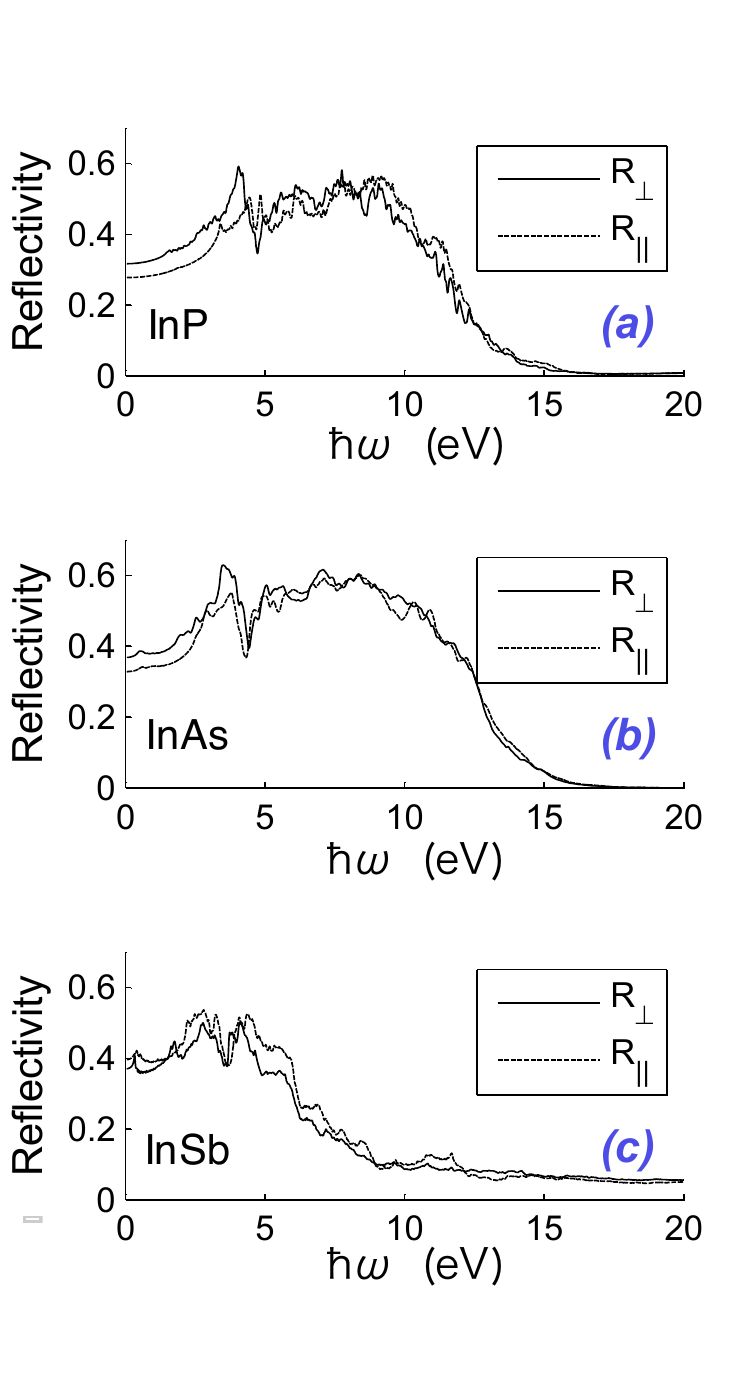}
\caption{ Calculated reflectivity spectra at normal incidence for {\bf(a)} InP, {\bf(b)} InAs and {\bf(c)} InSb }
\label{fig:R:InPAsSb}
\end{figure}

\begin{sidewaystable}
{\footnotesize
\begin{tabular}{|c|c|c c c c c|c c c c c|c c c c c|c c c c c|}
\cline{3-22}
\multicolumn{2}{c|}{} & \multicolumn{5}{c|}{for $\epsilon^{\prime}_{\perp}$} & \multicolumn{5}{c|}{for $\epsilon^{\prime\prime}_{\perp}$} & \multicolumn{5}{c|}{for $\epsilon^{\prime}_{||}$} & \multicolumn{5}{c|}{for $\epsilon^{\prime\prime}_{||}$} \\
\hline
material & $\hbar\omega_K$ (eV) & $f$ & $\Omega_1$ & $\Gamma_1$ & $\Omega_2$ & $\Gamma_2$ & $f$ & $\Omega_1$ & $\Gamma_1$ & $\Omega_2$ & $\Gamma_2$ & $f$ & $\Omega_1$ & $\Gamma_1$ & $\Omega_2$ & $\Gamma_2$& $f$ & $\Omega_1$ & $\Gamma_1$ & $\Omega_2$ & $\Gamma_2$\\\hline
AlP&4&130.1&5.639&6.064&4.927&-0.7225&1.062&2.075&-9.482$\times10^{-6}$&4.059&0.2963&137.3&6.702&7.095&5.742&8.292$\times10^{-6}$&0.3597&2.045&-4.889$\times10^{-6}$&4.374&0.2395\\
AlAs&3.5&98.66&5.281&7.39&3.945&-0.7937&6.019&1.775&-1.505$\times10^{-5}$&3.525&0.3752&111&5.818&5.958&4.774&1.94$\times10^{-5}$&0.3517&1.775&-1.349$\times10^{-5}$&3.564&0.3594\\
AlSb&3.5&173.3&24.43&1048&4.006&-0.9518&29.01&1.775&-2.341$\times10^{-5}$&3.574&0.6396&123.5&6.17&8.954&4.006&-0.7481&10.49&1.775&-1.476$\times10^{-5}$&3.605&0.4174\\
GaP&4&214.4&13.87&104.8&4.715&-1.357&18.66&2.045&-2.168$\times10^{-5}$&4.011&0.6816&142.9&6.11&6.582&5.157&-3.88$\times10^{-6}$&5.39&2.045&-8.046$\times10^{-6}$&4.257&0.3438\\
GaAs&2.5&35.06&2.279&1.548&2.781&0.3765&6.239&1.295&-0.0002109&2.064&0.9754&160.6&22.74&666.6&3.696&1.721&2.93&1.295&-5.471$\times10^{-5}$&2.515&0.6727\\
GaSb&1.5&21.1&1.652&1.416&1.825&0.3051&109.5&0.7851&-3.005$\times10^{-6}$&2.201&0.0881&46.47&2.319&2.244&2.424&0.0002&0.4912&0.7852&-0.0001&1.488&0.6783\\
InP&2.5&43.84&2.616&1.958&2.89&0.4301&14.99&1.295&-4.324$\times10^{-5}$&2.962&1.289&74.09&3.942&3.117&4.005&-9.183$\times10^{-5}$&1.076&1.295&-6.656$\times10^{-5}$&2.502&0.8105\\
InAs&1&10.68&1.079&0.9922&1.269&0.2729&2.157&0.5467&7.456$\times10^{-5}$&1.042&0.7514&9.755&1.186&1.171&1.319&-0.3299&1.332&0.5424&-5.915$\times10^{-5}$&1.17&0.7103\\
InSb&0.5&2.54&0.5128&0.429&0.6435&-0.1199&0.3746&0.2754&-0.0012&0.414&0.2541&3.401&0.5733&0.4854&0.6665&-0.1552&0.7352&0.2748&-0.0003&0.5696&0.2978\\
\hline
\end{tabular}
}
\caption{Fit coefficients for the dielectric constants for the III-V WZ semiconductors about their respective band gaps. $\hbar\omega_K$ is the upper cutoff energy up until which these fits are valid.}
\label{tab:WZ-epsfit}
\end{sidewaystable}

For WZ materials, with the inclusion of spin-orbit interactions, all zone center states belong to either $\Gamma_7$, $\Gamma_8$ or $\Gamma_9$ representations (all in double group notation).
The zone center conduction band minima have either $\Gamma_7$ or $\Gamma_8$ symmetry and the valance band states are the $\Gamma_9$ heavy-hole, $\Gamma_7^1$ light-hole and a $\Gamma_7$ split-off-hole.
The interband transitions for WZ are dictated by the polarization dependent optical selection rules \cite{Birman1959b,Streitwolf1969}. For light polarized parallel to the $c$-axis ($E_{||}$), only transitions between states with the same symmetry are allowed, {\it i.e.} $\Gamma_7 \leftrightarrow \Gamma_7$, $\Gamma_8 \leftrightarrow \Gamma_8$ and $\Gamma_9 \leftrightarrow \Gamma_9$.  For light polarized perpendicular to the $c$-axis($E_{\perp}$), the allowed transitions are $\Gamma_7 \leftrightarrow \Gamma_7$, $\Gamma_8 \leftrightarrow \Gamma_8$ , $\Gamma_9 \leftrightarrow \Gamma_7$ and $\Gamma_9 \leftrightarrow \Gamma_8$. Note, that the $\Gamma_7 \leftrightarrow \Gamma_8$ transition is forbidden for all polarizations.

The other high symmetry k-point transitions  allowed for $E_{||}$ are $A_{7,8}\leftrightarrow A_{7,8}$, $A_9\leftrightarrow A_9$, $K_{4,5}\leftrightarrow K_{4,5}$, $K_{6}\leftrightarrow K_{6}$, $H_{4,5}\leftrightarrow H_{4,5}$ and $H_{6}\leftrightarrow H_{6}$.
 For $E_{\perp}$ the allowed dipole transitions are $A_{7,8}\leftrightarrow A_{7,8}$, $A_9\leftrightarrow A_{7,8}$, $K_{4,5}\leftrightarrow K_{6}$ and $H_{4,5}\leftrightarrow H_{6}$. All M and U valley transitions ({\it i.e.} $M_{5}\leftrightarrow M_{5}$ and $L_{5}\leftrightarrow L_{5}$) are allowed for all polarizations. For a comprehensive list of optical selection rules for WZ at various high symmetry points and along various directions, see Refs. \onlinecite{Streitwolf1969} and \onlinecite{Mojumder1982} .

The zone center transition energies for the first thirteen states and their respective irreducible representations are listed in Ref.\onlinecite{De2010prb}. Our band structure calculations show that the indirect gap ZB semiconductors (AlP, AlAs, AlSb and GaP) have direct band gaps in WZ phase \cite{De2010prb} with $\Gamma_8$ conduction band minima.
These materials will be partially optically active, with transitions from the $\Gamma_9$ heavy-hole bands allowed for $E_{\perp}$, making them potentially technologically important.
%With direct optical transitions from the $\Gamma_9$ heavy-hole bands for $E_{\perp}$, these partially active large bandgap WZ semiconductors could be technologically important for optics.
The other semiconductors with $\Gamma_8$ conduction band minima in WZ phase are GaAs and GaSb.
The Indium containing WZ semiconductors all have direct band gaps  with $\Gamma_7$ conduction band minima and are optically active for all polarizations of the electric field.

The calculated $\epsilon^{\prime}(\omega)$ and $\epsilon^{\prime\prime}(\omega)$ for light polarized parallel and perpendicular to the $c$-axis for the nine III-V semiconductors of interest are shown in Figs.~\ref{fig:eps:AlPAsSb}-\ref{fig:eps:InPAsSb} The corresponding reflectivity spectra for both polarizations is shown in Figs.~\ref{fig:R:AlPAsSb}-\ref{fig:eps:InPAsSb}. The reflectivity spectra for the different WZ crystals are distinct and depend on the details of its electronic band structure. For the WZ semiconductors, the most prominent features are typically seen up to about $7~\rm eV$. As expected, all the WZ phase semiconductors exhibit optical anisotropy, the degree of which varies with the semiconductor. The reflectivity spectra for the two different polarizations shows several peaks which originate from interband transitions along various high symmetry points.

In order to illustrate the the dielectric function's variations about the fundamental absorption edge (FAE), we fit the numerically calculated $\epsilon_{\perp}$ and $\epsilon_{||}$ (in the vicinity of their respective band gaps) to the Lorentz oscillator model. The real and imaginary parts of which are
\begin{eqnarray}
\epsilon^{\prime}(\omega)=1-f\displaystyle\sum_{j=1}^{2}\frac{\omega^2-\Omega_j^2}{(\omega^2-\Omega_j^2)^2-(\Gamma_j\omega)^2}
\label{Lorentz_eps1}
\end{eqnarray}
\begin{eqnarray}
\epsilon^{\prime\prime}(\omega)=f\displaystyle\sum_{j=1}^{2}\frac{\Gamma_j\omega}{(\omega^2-\Omega_j^2)^2-(\Gamma_j\omega)^2}
\label{Lorentz_eps2}
\end{eqnarray}
where, $f,~\Omega_j$ and $\Gamma_j$ were used as fitting parameters and are listed in table \ref{tab:WZ-epsfit}. Note that the real and imaginary parts of the dielectric functions were separately fit to Eq.\ref{Lorentz_eps1} and Eq.\ref{Lorentz_eps2}.

It is seen, that in general, $\epsilon^{\prime\prime}_{\perp}$ is greater than $\epsilon^{\prime\prime}_{||}$. Based on optical selection rules, this can be explained by the fact that for $E_{\perp}$ a transition between the $\Gamma_9$ heavy-hole (HH) and the $\Gamma_7$ (or $\Gamma_8$) conduction band (CB) is allowed.
Hence the FAE is at the bandgap ($E_g$).  For $E_{||}$ the allowed transition with the lowest energy will be between the $\Gamma_7$ light-hole (LH) and $\Gamma_7$.
As a result for $E_{||}$, InP, InAs and InSb, the FAE will be at $E_g+\Delta_1$ where $\Delta_1$ is the energy difference between the $\Gamma_9$ heavy-hole and $\Gamma_7$ light-hole. In general, Indium containing compounds have prominent features in the reflectivity spectra near the FAE as a number of valance band to conduction band transitions are allowed. In all others (AlP, AlAs, AlSb, GaP, GaAs and GaSb) no noticeable sharp features are seen in the vicinity of the FAE since they all have $\Gamma_8$ CBs and the optical selection rules forbid transitions from the LH and the split-off hole for all polarizations and from the HH for $E_{||}$.

The spin-orbit coupling can alter the ordering of the valence band states in different WZ semiconductors. In our calculations, the valance band states in all materials except InSb  have normal ordering ({\it i.e} $\Gamma_9$, $\Gamma_7$, $\Gamma_7$ \cite{Birman1959b,Thomas1959}). The ordering of the valence band states in InSb is complicated by the very large spin-orbit splitting which forces the $\Gamma_7$ split-off hole bellow the next $\Gamma_9$ state, resulting in the unusual $\Gamma_9$, $\Gamma_7$, $\Gamma_9$, $\Gamma_7$ valence band ordering. Attempts to optically measure the spin-orbit splitting energy in WZ type InSb could be complicated by this.

%=================================================================================================================================================
\section{Summary \label{sec:summary}}
%=================================================================================================================================================
In summary we have calculated the optical properties of AlP, AlAs, AlSb, GaP, GaAs, GaSb, InP, InAs and InSb in WZ phase using empirical pseudopotentials. Their complex dielectric function was evaluated as a function of incident photon energy up to $20~\rm eV$ for light polarized parallel and perpendicular to the $c$-axis. The dielectric functions were calculated in the dipole approximation by evaluating the optical momentum matrix elements using empirical pseudopotential wave functions (from Ref.\onlinecite{De2010prb}) on a grid of about $4.5\times 10^4$ special k-points. Corrections to the pseudo-momentum-matrix elements to account for the missing core states are introduced via a scaling factor, which is determined from the ratio of the calculated to measured static dielectric function for the corresponding zincblende polytypes.
The reflectivity spectra for all nine WZ semiconductors was also calculated for both polarizations and was seen to exhibit optical anisotropy as well. We have also calculated the static frequency dielectric constants.
% for parallel and perpendicular polarizations of the electric field for the nine III-V WZ phase semiconductors.


\begin{thebibliography}
\expandafter\ifx\csname natexlab\endcsname\relax\def\natexlab#1{#1}\fi
\expandafter\ifx\csname bibnamefont\endcsname\relax
  \def\bibnamefont#1{#1}\fi
\expandafter\ifx\csname bibfnamefont\endcsname\relax
  \def\bibfnamefont#1{#1}\fi
\expandafter\ifx\csname citenamefont\endcsname\relax
  \def\citenamefont#1{#1}\fi
\expandafter\ifx\csname url\endcsname\relax
  \def\url#1{\texttt{#1}}\fi
\expandafter\ifx\csname urlprefix\endcsname\relax\def\urlprefix{URL }\fi
\providecommand{\bibinfo}[2]{#2}
\providecommand{\eprint}[2][]{\url{#2}}

\bibitem[{\citenamefont{Cardona and Harbeke}(1965)}]{Cardona1965}
\bibinfo{author}{\bibfnamefont{M.}~\bibnamefont{Cardona}} \bibnamefont{and}
  \bibinfo{author}{\bibfnamefont{G.}~\bibnamefont{Harbeke}},
  \bibinfo{journal}{Phys. Rev.} \textbf{\bibinfo{volume}{137}},
  \bibinfo{pages}{A1467} (\bibinfo{year}{1965}).

\bibitem[{\citenamefont{Reynolds et~al.}(1965)\citenamefont{Reynolds, Litton,
  and Collins}}]{Reynolds1965}
\bibinfo{author}{\bibfnamefont{D.~C.} \bibnamefont{Reynolds}},
  \bibinfo{author}{\bibfnamefont{C.~W.} \bibnamefont{Litton}},
  \bibnamefont{and} \bibinfo{author}{\bibfnamefont{T.~C.}
  \bibnamefont{Collins}}, \bibinfo{journal}{Phys. Rev.}
  \textbf{\bibinfo{volume}{140}}, \bibinfo{pages}{A1726}
  (\bibinfo{year}{1965}).

\bibitem[{\citenamefont{Xu and Ching}(1993)}]{Xu1993}
\bibinfo{author}{\bibfnamefont{Y.-N.} \bibnamefont{Xu}} \bibnamefont{and}
  \bibinfo{author}{\bibfnamefont{W.~Y.} \bibnamefont{Ching}},
  \bibinfo{journal}{Phys. Rev. B} \textbf{\bibinfo{volume}{48}},
  \bibinfo{pages}{4335} (\bibinfo{year}{1993}).

\bibitem[{\citenamefont{Ninomiya and Adachi}(1995)}]{Ninomiya1995}
\bibinfo{author}{\bibfnamefont{S.}~\bibnamefont{Ninomiya}} \bibnamefont{and}
  \bibinfo{author}{\bibfnamefont{S.}~\bibnamefont{Adachi}},
  \bibinfo{journal}{Journal of Applied Physics} \textbf{\bibinfo{volume}{78}},
  \bibinfo{pages}{1183} (\bibinfo{year}{1995}).

\bibitem[{\citenamefont{Kawashima et~al.}(1997)\citenamefont{Kawashima,
  Yoshikawa, Adachi, Fuke, and Ohtsuka}}]{Kawashima1997}
\bibinfo{author}{\bibfnamefont{T.}~\bibnamefont{Kawashima}},
  \bibinfo{author}{\bibfnamefont{H.}~\bibnamefont{Yoshikawa}},
  \bibinfo{author}{\bibfnamefont{S.}~\bibnamefont{Adachi}},
  \bibinfo{author}{\bibfnamefont{S.}~\bibnamefont{Fuke}}, \bibnamefont{and}
  \bibinfo{author}{\bibfnamefont{K.}~\bibnamefont{Ohtsuka}},
  \bibinfo{journal}{Journal of Applied Physics} \textbf{\bibinfo{volume}{82}},
  \bibinfo{pages}{3528} (\bibinfo{year}{1997}).

\bibitem[{\citenamefont{Alemu et~al.}(1998)\citenamefont{Alemu, Gil, Julier,
  and Nakamura}}]{Alemu1998}
\bibinfo{author}{\bibfnamefont{A.}~\bibnamefont{Alemu}},
  \bibinfo{author}{\bibfnamefont{B.}~\bibnamefont{Gil}},
  \bibinfo{author}{\bibfnamefont{M.}~\bibnamefont{Julier}}, \bibnamefont{and}
  \bibinfo{author}{\bibfnamefont{S.}~\bibnamefont{Nakamura}},
  \bibinfo{journal}{Phys. Rev. B} \textbf{\bibinfo{volume}{57}},
  \bibinfo{pages}{3761} (\bibinfo{year}{1998}).

\bibitem[{\citenamefont{Birman}(1959{\natexlab{a}})}]{Birman1959b}
\bibinfo{author}{\bibfnamefont{J.~L.} \bibnamefont{Birman}},
  \bibinfo{journal}{Phys. Rev.} \textbf{\bibinfo{volume}{114}},
  \bibinfo{pages}{1490} (\bibinfo{year}{1959}{\natexlab{a}}).

\bibitem[{\citenamefont{Streitwolf}(1969)}]{Streitwolf1969}
\bibinfo{author}{\bibfnamefont{A.~H.~W.} \bibnamefont{Streitwolf}},
  \bibinfo{journal}{Physica Status Solidi (b)} \textbf{\bibinfo{volume}{33}},
  \bibinfo{pages}{225} (\bibinfo{year}{1969}).

\bibitem[{\citenamefont{Wang et~al.}(2001)\citenamefont{Wang, Gudiksen, Duan,
  Cui, and Lieber}}]{Wang2001}
\bibinfo{author}{\bibfnamefont{J.}~\bibnamefont{Wang}},
  \bibinfo{author}{\bibfnamefont{M.~S.} \bibnamefont{Gudiksen}},
  \bibinfo{author}{\bibfnamefont{X.}~\bibnamefont{Duan}},
  \bibinfo{author}{\bibfnamefont{Y.}~\bibnamefont{Cui}}, \bibnamefont{and}
  \bibinfo{author}{\bibfnamefont{C.~M.} \bibnamefont{Lieber}},
  \bibinfo{journal}{Science} \textbf{\bibinfo{volume}{293}},
  \bibinfo{pages}{1455} (\bibinfo{year}{2001}).

\bibitem[{\citenamefont{Mattila et~al.}(2006)\citenamefont{Mattila,
  Hakkarainen, Mulot, and Lipsanen}}]{Mattila2006}
\bibinfo{author}{\bibfnamefont{M.}~\bibnamefont{Mattila}},
  \bibinfo{author}{\bibfnamefont{T.}~\bibnamefont{Hakkarainen}},
  \bibinfo{author}{\bibfnamefont{M.}~\bibnamefont{Mulot}}, \bibnamefont{and}
  \bibinfo{author}{\bibfnamefont{H.}~\bibnamefont{Lipsanen}},
  \bibinfo{journal}{Nanotechnology} \textbf{\bibinfo{volume}{17}},
  \bibinfo{pages}{1580} (\bibinfo{year}{2006}).

\bibitem[{\citenamefont{Mattila et~al.}(2007)\citenamefont{Mattila,
  Hakkarainen, Lipsanen, Jiang, and Kauppinen}}]{Mattila2007}
\bibinfo{author}{\bibfnamefont{M.}~\bibnamefont{Mattila}},
  \bibinfo{author}{\bibfnamefont{T.}~\bibnamefont{Hakkarainen}},
  \bibinfo{author}{\bibfnamefont{H.}~\bibnamefont{Lipsanen}},
  \bibinfo{author}{\bibfnamefont{H.}~\bibnamefont{Jiang}}, \bibnamefont{and}
  \bibinfo{author}{\bibfnamefont{E.~I.} \bibnamefont{Kauppinen}},
  \bibinfo{journal}{Appl. Phys. Lett.} \textbf{\bibinfo{volume}{90}},
  \bibinfo{eid}{033101} (\bibinfo{year}{2007}).

\bibitem[{\citenamefont{Mishra et~al.}(2007)\citenamefont{Mishra, Titova,
  Hoang, Jackson, Smith, Yarrison-Rice, Kim, Joyce, Gao, Tan
  et~al.}}]{Mishra2007}
\bibinfo{author}{\bibfnamefont{A.}~\bibnamefont{Mishra}},
  \bibinfo{author}{\bibfnamefont{L.~V.} \bibnamefont{Titova}},
  \bibinfo{author}{\bibfnamefont{T.~B.} \bibnamefont{Hoang}},
  \bibinfo{author}{\bibfnamefont{H.~E.} \bibnamefont{Jackson}},
  \bibinfo{author}{\bibfnamefont{L.~M.} \bibnamefont{Smith}},
  \bibinfo{author}{\bibfnamefont{J.~M.} \bibnamefont{Yarrison-Rice}},
  \bibinfo{author}{\bibfnamefont{Y.}~\bibnamefont{Kim}},
  \bibinfo{author}{\bibfnamefont{H.~J.} \bibnamefont{Joyce}},
  \bibinfo{author}{\bibfnamefont{Q.}~\bibnamefont{Gao}},
  \bibinfo{author}{\bibfnamefont{H.~H.} \bibnamefont{Tan}},
  \bibnamefont{et~al.}, \bibinfo{journal}{Appl. Phys. Lett.}
  \textbf{\bibinfo{volume}{91}}, \bibinfo{eid}{263104}
  (pages~\bibinfo{numpages}{3}) (\bibinfo{year}{2007}).

\bibitem[{\citenamefont{Lan et~al.}(2008)\citenamefont{Lan, Giblin, Protasenko,
  and Kuno}}]{Lan2008}
\bibinfo{author}{\bibfnamefont{A.}~\bibnamefont{Lan}},
  \bibinfo{author}{\bibfnamefont{J.}~\bibnamefont{Giblin}},
  \bibinfo{author}{\bibfnamefont{V.}~\bibnamefont{Protasenko}},
  \bibnamefont{and} \bibinfo{author}{\bibfnamefont{M.}~\bibnamefont{Kuno}},
  \bibinfo{journal}{Applied Physics Letters} \textbf{\bibinfo{volume}{92}},
  \bibinfo{eid}{183110} (pages~\bibinfo{numpages}{3}) (\bibinfo{year}{2008}).

\bibitem[{\citenamefont{Kobayashi et~al.}(2008)\citenamefont{Kobayashi, Fukui,
  Motohisa, and Fukui}}]{Kobayashi2008}
\bibinfo{author}{\bibfnamefont{Y.}~\bibnamefont{Kobayashi}},
  \bibinfo{author}{\bibfnamefont{M.}~\bibnamefont{Fukui}},
  \bibinfo{author}{\bibfnamefont{J.}~\bibnamefont{Motohisa}}, \bibnamefont{and}
  \bibinfo{author}{\bibfnamefont{T.}~\bibnamefont{Fukui}},
  \bibinfo{journal}{Physica E: Low-dimensional Systems and Nanostructures}
  \textbf{\bibinfo{volume}{40}}, \bibinfo{pages}{2204 } (\bibinfo{year}{2008}),
  ISSN \bibinfo{issn}{1386-9477}, \bibinfo{note}{13th International Conference
  on Modulated Semiconductor Structures}.

\bibitem[{\citenamefont{Novikov et~al.}(2010)\citenamefont{Novikov, Serov,
  Filosofov, Shtrom, Talalaev, Vyvenko, Ubyivovk, Samsonenko, Bouravleuv,
  Soshnikov et~al.}}]{Novikov2010}
\bibinfo{author}{\bibfnamefont{B.~V.} \bibnamefont{Novikov}},
  \bibinfo{author}{\bibfnamefont{S.~Y.} \bibnamefont{Serov}},
  \bibinfo{author}{\bibfnamefont{N.~G.} \bibnamefont{Filosofov}},
  \bibinfo{author}{\bibfnamefont{I.~V.} \bibnamefont{Shtrom}},
  \bibinfo{author}{\bibfnamefont{V.~G.} \bibnamefont{Talalaev}},
  \bibinfo{author}{\bibfnamefont{O.~F.} \bibnamefont{Vyvenko}},
  \bibinfo{author}{\bibfnamefont{E.~V.} \bibnamefont{Ubyivovk}},
  \bibinfo{author}{\bibfnamefont{Y.~B.} \bibnamefont{Samsonenko}},
  \bibinfo{author}{\bibfnamefont{A.~D.} \bibnamefont{Bouravleuv}},
  \bibinfo{author}{\bibfnamefont{I.~P.} \bibnamefont{Soshnikov}},
  \bibnamefont{et~al.}, \bibinfo{journal}{physica status solidi (RRL) – Rapid
  Research Letters} \textbf{\bibinfo{volume}{4}}, \bibinfo{pages}{175–}
  (\bibinfo{year}{2010}).

\bibitem[{\citenamefont{Sercel and Vahala}(1990)}]{Sercel:1990p646}
\bibinfo{author}{\bibfnamefont{P.}~\bibnamefont{Sercel}} \bibnamefont{and}
  \bibinfo{author}{\bibfnamefont{K.}~\bibnamefont{Vahala}},
  \bibinfo{journal}{Physical Review B} \textbf{\bibinfo{volume}{42}},
  \bibinfo{pages}{3690} (\bibinfo{year}{1990}).

\bibitem[{\citenamefont{Persson and Xu}(2004)}]{Persson2004}
\bibinfo{author}{\bibfnamefont{M.~P.} \bibnamefont{Persson}} \bibnamefont{and}
  \bibinfo{author}{\bibfnamefont{H.~Q.} \bibnamefont{Xu}},
  \bibinfo{journal}{Phys. Rev. B} \textbf{\bibinfo{volume}{70}},
  \bibinfo{pages}{161310} (\bibinfo{year}{2004}).

\bibitem[{\citenamefont{De and Pryor}(2010)}]{De2010prb}
\bibinfo{author}{\bibfnamefont{A.}~\bibnamefont{De}} \bibnamefont{and}
  \bibinfo{author}{\bibfnamefont{C.~E.} \bibnamefont{Pryor}},
  \bibinfo{journal}{Phys. Rev. B} \textbf{\bibinfo{volume}{81}},
  \bibinfo{pages}{155210} (\bibinfo{year}{2010}).

\bibitem[{\citenamefont{Birman}(1959{\natexlab{b}})}]{Birman.1959}
\bibinfo{author}{\bibfnamefont{J.~L.} \bibnamefont{Birman}},
  \bibinfo{journal}{Phys. Rev. Lett.} \textbf{\bibinfo{volume}{2}},
  \bibinfo{pages}{157} (\bibinfo{year}{1959}{\natexlab{b}}).

\bibitem[{\citenamefont{Joannopoulos and Cohen}(1973)}]{Joannopoulos1973}
\bibinfo{author}{\bibfnamefont{J.~D.} \bibnamefont{Joannopoulos}}
  \bibnamefont{and} \bibinfo{author}{\bibfnamefont{M.~L.} \bibnamefont{Cohen}},
  \bibinfo{journal}{Phys. Rev. B} \textbf{\bibinfo{volume}{8}},
  \bibinfo{pages}{2733} (\bibinfo{year}{1973}).

\bibitem[{\citenamefont{Pugh et~al.}(1999)\citenamefont{Pugh, Dugdale, Brand,
  and Abram}}]{Pugh1999}
\bibinfo{author}{\bibfnamefont{S.~K.} \bibnamefont{Pugh}},
  \bibinfo{author}{\bibfnamefont{D.~J.} \bibnamefont{Dugdale}},
  \bibinfo{author}{\bibfnamefont{S.}~\bibnamefont{Brand}}, \bibnamefont{and}
  \bibinfo{author}{\bibfnamefont{R.~A.} \bibnamefont{Abram}},
  \bibinfo{journal}{J. Appl. Phys.} \textbf{\bibinfo{volume}{86}},
  \bibinfo{pages}{3768} (\bibinfo{year}{1999}).

\bibitem[{\citenamefont{Pennington and Goldsman}(2001)}]{Pennington2001}
\bibinfo{author}{\bibfnamefont{G.}~\bibnamefont{Pennington}} \bibnamefont{and}
  \bibinfo{author}{\bibfnamefont{N.}~\bibnamefont{Goldsman}},
  \bibinfo{journal}{Phys. Rev. B} \textbf{\bibinfo{volume}{64}},
  \bibinfo{pages}{045104} (\bibinfo{year}{2001}).

\bibitem[{\citenamefont{Fritsch et~al.}(2006)\citenamefont{Fritsch, Schmidt,
  and Grundmann}}]{Fritsch2006}
\bibinfo{author}{\bibfnamefont{D.}~\bibnamefont{Fritsch}},
  \bibinfo{author}{\bibfnamefont{H.}~\bibnamefont{Schmidt}}, \bibnamefont{and}
  \bibinfo{author}{\bibfnamefont{M.}~\bibnamefont{Grundmann}},
  \bibinfo{journal}{Appl. Phys. Lett.} \textbf{\bibinfo{volume}{88}},
  \bibinfo{eid}{134104} (\bibinfo{year}{2006}).

\bibitem[{\citenamefont{Cohen and Chelikowsky}(1988)}]{Cohen.book}
\bibinfo{author}{\bibfnamefont{M.}~\bibnamefont{Cohen}} \bibnamefont{and}
  \bibinfo{author}{\bibfnamefont{J.~R.} \bibnamefont{Chelikowsky}},
  \emph{\bibinfo{title}{Electronic Structure and Optical Properties of
  Semiconductors}} (\bibinfo{publisher}{Springer}, \bibinfo{address}{Berlin},
  \bibinfo{year}{1988}).

\bibitem[{\citenamefont{Kageshima and Shiraishi}(1997)}]{Kageshima1997}
\bibinfo{author}{\bibfnamefont{H.}~\bibnamefont{Kageshima}} \bibnamefont{and}
  \bibinfo{author}{\bibfnamefont{K.}~\bibnamefont{Shiraishi}},
  \bibinfo{journal}{Phys. Rev. B} \textbf{\bibinfo{volume}{56}},
  \bibinfo{pages}{14985} (\bibinfo{year}{1997}).

\bibitem[{\citenamefont{Adolph et~al.}(2001)\citenamefont{Adolph,
  Furthm\"uller, and Bechstedt}}]{Adolph2001}
\bibinfo{author}{\bibfnamefont{B.}~\bibnamefont{Adolph}},
  \bibinfo{author}{\bibfnamefont{J.}~\bibnamefont{Furthm\"uller}},
  \bibnamefont{and}
  \bibinfo{author}{\bibfnamefont{F.}~\bibnamefont{Bechstedt}},
  \bibinfo{journal}{Phys. Rev. B} \textbf{\bibinfo{volume}{63}},
  \bibinfo{pages}{125108} (\bibinfo{year}{2001}).

\bibitem[{\citenamefont{Pickard and Mauri}(2001)}]{Pickard2001}
\bibinfo{author}{\bibfnamefont{C.~J.} \bibnamefont{Pickard}} \bibnamefont{and}
  \bibinfo{author}{\bibfnamefont{F.}~\bibnamefont{Mauri}},
  \bibinfo{journal}{Phys. Rev. B} \textbf{\bibinfo{volume}{63}},
  \bibinfo{pages}{245101} (\bibinfo{year}{2001}).

\bibitem[{\citenamefont{Monachesi et~al.}(2001)\citenamefont{Monachesi, Marini,
  andM. Palummo, and Sole}}]{Monachesi2001}
\bibinfo{author}{\bibfnamefont{P.}~\bibnamefont{Monachesi}},
  \bibinfo{author}{\bibfnamefont{A.}~\bibnamefont{Marini}},
  \bibinfo{author}{\bibfnamefont{G.~O.} \bibnamefont{andM. Palummo}},
  \bibnamefont{and} \bibinfo{author}{\bibfnamefont{R.~D.} \bibnamefont{Sole}},
  \bibinfo{journal}{physica status solidi (a)} \textbf{\bibinfo{volume}{184}},
  \bibinfo{pages}{101} (\bibinfo{year}{2001}).

\bibitem[{\citenamefont{L.~P.~Mosteller and Wooten}(1968)}]{Mosteller1968}
\bibinfo{author}{\bibfnamefont{J.}~\bibnamefont{L.~P.~Mosteller}}
  \bibnamefont{and} \bibinfo{author}{\bibfnamefont{F.}~\bibnamefont{Wooten}},
  \bibinfo{journal}{J. Opt. Soc. Am.} \textbf{\bibinfo{volume}{58}},
  \bibinfo{pages}{511} (\bibinfo{year}{1968}).

\bibitem[{\citenamefont{Brehat and Wyncke}(1991)}]{Brehat1991}
\bibinfo{author}{\bibfnamefont{F.}~\bibnamefont{Brehat}} \bibnamefont{and}
  \bibinfo{author}{\bibfnamefont{B.}~\bibnamefont{Wyncke}},
  \bibinfo{journal}{Journal of Physics D: Applied Physics}
  \textbf{\bibinfo{volume}{24}}, \bibinfo{pages}{2055} (\bibinfo{year}{1991}).

\bibitem[{\citenamefont{Wang and Klein}(1981)}]{Wang1981}
\bibinfo{author}{\bibfnamefont{C.~S.} \bibnamefont{Wang}} \bibnamefont{and}
  \bibinfo{author}{\bibfnamefont{B.~M.} \bibnamefont{Klein}},
  \bibinfo{journal}{Phys. Rev. B} \textbf{\bibinfo{volume}{24}},
  \bibinfo{pages}{3417} (\bibinfo{year}{1981}).

\bibitem[{\citenamefont{Monkhorst and Pack}(1976)}]{Monkhorst1976}
\bibinfo{author}{\bibfnamefont{H.~J.} \bibnamefont{Monkhorst}}
  \bibnamefont{and} \bibinfo{author}{\bibfnamefont{J.~D.} \bibnamefont{Pack}},
  \bibinfo{journal}{Phys. Rev. B} \textbf{\bibinfo{volume}{13}},
  \bibinfo{pages}{5188} (\bibinfo{year}{1976}).

\bibitem[{\citenamefont{Read and Needs}(1991)}]{Read1991}
\bibinfo{author}{\bibfnamefont{A.~J.} \bibnamefont{Read}} \bibnamefont{and}
  \bibinfo{author}{\bibfnamefont{R.~J.} \bibnamefont{Needs}},
  \bibinfo{journal}{Phys. Rev. B} \textbf{\bibinfo{volume}{44}},
  \bibinfo{pages}{13071} (\bibinfo{year}{1991}).

\bibitem[{\citenamefont{Adolph et~al.}(1996)\citenamefont{Adolph, Gavrilenko,
  Tenelsen, Bechstedt, and Del~Sole}}]{Adolph1996}
\bibinfo{author}{\bibfnamefont{B.}~\bibnamefont{Adolph}},
  \bibinfo{author}{\bibfnamefont{V.~I.} \bibnamefont{Gavrilenko}},
  \bibinfo{author}{\bibfnamefont{K.}~\bibnamefont{Tenelsen}},
  \bibinfo{author}{\bibfnamefont{F.}~\bibnamefont{Bechstedt}},
  \bibnamefont{and} \bibinfo{author}{\bibfnamefont{R.}~\bibnamefont{Del~Sole}},
  \bibinfo{journal}{Phys. Rev. B} \textbf{\bibinfo{volume}{53}},
  \bibinfo{pages}{9797} (\bibinfo{year}{1996}).

\bibitem[{\citenamefont{Madelung}(2004)}]{Madelung}
\bibinfo{editor}{\bibfnamefont{O.}~\bibnamefont{Madelung}}, ed.,
  \emph{\bibinfo{title}{Semiconductors Data Handbook}}
  (\bibinfo{publisher}{Springer-Verlag}, \bibinfo{address}{Berlin Heidelberg
  New York}, \bibinfo{year}{2004}), \bibinfo{edition}{$3^{rd}$} ed.

\bibitem[{\citenamefont{Palik}(1998)}]{Palik.Handbook}
\bibinfo{editor}{\bibfnamefont{E.~D.} \bibnamefont{Palik}}, ed.,
  \emph{\bibinfo{title}{Handbook of optical constants of solids}}
  (\bibinfo{publisher}{Academic Press}, \bibinfo{address}{London},
  \bibinfo{year}{1998}).

\bibitem[{\citenamefont{Mojumder}(1982)}]{Mojumder1982}
\bibinfo{author}{\bibfnamefont{M.~A.} \bibnamefont{Mojumder}},
  \bibinfo{journal}{Solid State Communications} \textbf{\bibinfo{volume}{43}},
  \bibinfo{pages}{13 } (\bibinfo{year}{1982}), ISSN \bibinfo{issn}{0038-1098}.

\bibitem[{\citenamefont{Thomas and Hopfield}(1959)}]{Thomas1959}
\bibinfo{author}{\bibfnamefont{D.~G.} \bibnamefont{Thomas}} \bibnamefont{and}
  \bibinfo{author}{\bibfnamefont{J.~J.} \bibnamefont{Hopfield}},
  \bibinfo{journal}{Phys. Rev.} \textbf{\bibinfo{volume}{116}},
  \bibinfo{pages}{573} (\bibinfo{year}{1959}).

\end{thebibliography}
\end{document}